\documentclass[twocolumn,aps,prl,superscriptaddress,nofootinbib,showpacs]{revtex4-2}
\usepackage{graphicx}
\usepackage{epstopdf}
\usepackage{color}
\graphicspath{{graphs/}{fig/}}
\usepackage{amsmath}
\usepackage{amssymb}
\usepackage{bm}
\usepackage{mathrsfs}
\usepackage{slashed}
\usepackage{epsfig}
\usepackage{amsfonts}
\usepackage{hyperref}
\usepackage{bbm}
\usepackage{textcomp}
\usepackage{multirow}
\usepackage{float}
\usepackage{extarrows}

\newcommand{\sect}[1]{%
  \refstepcounter{section}%
  {\it\textbf{#1.} --- }%
}
\makeatletter\def\p@subsection{}\makeatother
\newcommand{\mi}{\mathrm{i}}
\newcommand{\md}{\mathrm{d}}

\newcommand{\cD}{{\mathcal{D}}}

\begin{document}
\title{Analyticity Bootstrap of Feynman Integrals}

\author{Xiang Li}
\email{lix-PHY@pku.edu.cn}
\affiliation{School of Physics and State Key Laboratory of Nuclear Physics and Technology, Peking University, Beijing 100871, China}

\author{Dao-Ming Mu}
\email{mudaoming@pku.edu.cn}
\affiliation{School of Physics and State Key Laboratory of Nuclear Physics and Technology, Peking University, Beijing 100871, China}

\author{Yan-Qing Ma}
\email{yqma@pku.edu.cn}
\affiliation{School of Physics and State Key Laboratory of Nuclear Physics and Technology, Peking University, Beijing 100871, China}
\affiliation{Center for High Energy Physics, Peking University, Beijing 100871, China}

\date{\today}
\begin{abstract}
We propose an analyticity bootstrap method to determine the reduction of Feynman integrals (FIs). The key observation is that, at any kinematic point, a dimensionally regularized FI possesses a Taylor expansion region (analytic region), and the finiteness of the Taylor region strongly constrains the integral reduction coefficients, which are rational functions of the kinematic variables. By writing the most general ansatz compatible with the singularity structure and then fixing the remaining parameters with only a few integration-by-parts (IBP) samples or asymptotic expansions, one can obtain the full reduction. We develop the method systematically in both single-variable and multi-variable cases, with many explicit examples. Interestingly, the differential equations (DEs) of the master integrals determine all these analyticity constraints and, conversely, the constraints facilitate the construction of the DEs. In particular, we show that the DEs in the auxiliary mass flow method can be constructed at significantly reduced cost and can even be fully determined without using IBP.
\end{abstract}

\maketitle
\allowdisplaybreaks

%%%%%%%%%%%%%%%%%%%%%%%%%%%%%%%%%%%%%%%%%%%%%%%%%

\sect{Introduction}
Feynman integrals (FIs) encode key information of quantum field theories, from precision phenomenology to fundamental theoretical explorations. Their computation is the core difficulty of precision calculations in both current collider physics and gravitational-wave physics (e.g.,~\cite{Heinrich:2020ybq, Bern:2019nnu,Abreu:2024fei,Henn:2025xrc,Becchetti:2025oyb,Driesse:2024feo,Driesse:2026qiz}). The mainstream strategy consists of two steps: first, reduce all FIs to a finite set of master integrals (MIs) using integration-by-parts (IBP) identities~\cite{Chetyrkin:1981qh,Laporta:2000dsw}; second, evaluate these MIs, most powerfully by solving the differential equations (DEs)~\cite{Kotikov:1990kg,Remiddi:1997ny,Gehrmann:1999as,Henn:2013pwa,Liu:2017jxz,Liu:2022chg}.

IBP reduction involves solving large linear systems over rational function fields, which becomes increasingly demanding in multiloop, multiscale problems. Finite-field reconstruction alleviates intermediate expression swell~\cite{vonManteuffel:2014ixa,Peraro:2016wsq,Peraro:2019svx,Smirnov:2019qkx,Klappert:2020nbg,Smirnov:2025prc}, while optimized IBP equations reduce the size of the systems to be solved~\cite{Gluza:2010ws,Schabinger:2011dz,vonManteuffel:2012np,Anastasiou:2004vj,Lee:2012cn,Lee:2013mka,Lange:2025fba,Wu:2023upw,Wu:2025aeg,Guan:2024byi,Chen:2024xwt}. Nevertheless, reconstructing the rational reduction coefficients can still require many numerical samples~\cite{Peraro:2016wsq,Peraro:2019svx,Klappert:2019emp}. The construction of the DEs also relies on integral reduction, since the derivatives must be reduced to the MIs. IBP reduction is thus the bottleneck of many state-of-the-art computations.

The analyticity of FIs has been used in various forms to constrain amplitudes and integrals, including amplitude bootstrap~\cite{Dixon:2011pw,Caron-Huot:2020bkp}, dimensional recurrence~\cite{Lee:2009dh}, singularity-based ansatz construction~\cite{Laurentis:2019bjh,DeLaurentis:2022otd,Boehm:2020ijp,Chawdhry:2023yyx,DeLaurentis:2025dxw}, correlations among functions~\cite{Liu:2023cgs,DeLaurentis:2023nss}, and integral bases~\cite{DeAngelis:2025agn,Song:2026xch}. Yet its full power has not been exploited. 

In this Letter, we show that a great deal of information is already contained in the analytic structure of FIs. FIs are not arbitrary functions: dimensionally regularized FIs possess Taylor series at any finite kinematic point within their analytic region. This structure imposes strong constraints on the rational functions in both the IBP reduction coefficients and the DE matrices, so that an ansatz with far fewer free parameters suffices and can be fixed with a minimal amount of additional input. We refer to it as the analyticity bootstrap (AB).

\sect{Setup}
\label{sec:setup}
Before proceeding, we recall the basic setup. A family of FIs is defined by
\begin{align}\label{eq:family}
{I}_{\vec{\nu}}=\int\left(\prod_{i=1}^{L}\frac{\md^{D}\ell_i}{\mi\pi^{D/2}}\right)
\frac{\cD_{K+1}^{-\nu_{K+1}}\cdots \cD_{N_\cD}^{-\nu_{N_\cD}}}
{\cD_1^{\nu_{1}}\cdots \cD_K^{\nu_{K}}},
\end{align}
where $L$ is the number of loops, $\ell_i$ are loop momenta, $D=4-2\epsilon$ is the spacetime dimension, $\cD_1,\ldots,\cD_K$ are inverse propagators, $\cD_{K+1},\ldots,\cD_{N_\cD}$ are irreducible scalar products, $N_\cD=L(L+1)/2 + LE$ and $E$ is the number of external momenta. A family forms a finite-dimensional linear space~\cite{Smirnov:2010hn}, so that any FI can be decomposed as
\begin{align}\label{eq:reduction}
F = \sum_{i=1}^{N} c_i\, I_i,\qquad c_i \in \mathbb{Q}(\epsilon, \vec{x}),
\end{align}
where $\{I_1,\ldots,I_N\}$ are MIs and $c_i$ are rational functions of the kinematic variables $\vec{x}$ (masses, Mandelstam invariants, etc.). The decomposition is usually achieved by using IBP identities.

We first consider a single kinematic variable $x$, with all other variables fixed. The MIs satisfy a system of first-order linear rational DEs,
\begin{align}\label{eq:de}
\partial_x \vec I = A(x)\,\vec I,\qquad A\in \mathrm{Mat}_{N\times N}\!\big(\mathbb{Q}(x)\big).
\end{align}
We denote the pole set of $A$ by $\mathcal{S}_A=\{\alpha_1,\alpha_2,\dots\}\subset\mathbb{C}$, excluding the pole at infinity. These poles include both physical and spurious singularities.

The Feynman parametric representation gives a local expansion of each MI near a singular point $\alpha\in\mathcal{S}_A$. Writing $z=x-\alpha$, its general form is~\cite{Bergere:1977wz,Friot:2005cu,Bogner:2010kv,Semenova:2018cwy,Lee:2017qql}
\begin{align}\label{eq:local}
I_i(x)=\sum_{\mu\in\Lambda_\alpha}\sum_{p=0}^{p_\mu}z^\mu(\log z)^p \sum_{n=0}^{\infty}b_{i,n}^{(\mu,p)}z^n+\sum_{n=0}^{\infty}a_{i,n}z^n,
\end{align}
where $\Lambda_\alpha$ is the set of characteristic exponents at $\alpha$, determined by the singular part of $A$, and $p_\mu\in\mathbb{Z}_{\geq 0}$ is the maximal logarithmic power associated with $\mu$. The first sum collects contributions from the \emph{non-analytic regions}, with non-integer powers or logarithms at general $\epsilon$, and the second sum is the \emph{analytic region} (Taylor series).

\sect{Analyticity constraints on reduction coefficients}
\label{sec:analyticity-constraints}
We now show how this local structure constrains the reduction. Our method rests on the following premises:

\textbf{(P1) Finiteness.} For each point $x=\alpha$, the MIs and $F$ have finite contributions from the analytic region, as is usually the case for dimensional regularization.\footnote{If necessary, the MIs and $F$ can be normalized by suitable powers of singular factors so that their contributions from the analytic region are finite at the relevant finite points. The DE matrix and reduction coefficients are then transformed consistently.}

\textbf{(P2) DEs.} The DE matrix $A(x)$ is taken as input. Its construction will be described in the next section.

\textbf{(P3) Poles.} For a minimal MI basis, the rational coefficients $c_i$ have poles only in $\mathcal{S}_A$. The MIs are analytic at an ordinary point $x=\beta\notin\mathcal{S}_A$ and linearly independent over $\mathbb{Q}$, so a pole of $c_i$ there would produce an uncancelled singularity, contradicting (P1). However, with redundant integrals, the coefficients may develop spurious poles outside $\mathcal{S}_A$.

By (P3), each $c_i$ is a rational function with poles confined to $\mathcal{S}_A$, so we can rewrite the reduction in Eq.~\eqref{eq:reduction} as
\begin{align}\label{eq:ansatz}
F = \sum_{\alpha\in\mathcal{S}_A} \frac{\vec{Q}^{(\alpha)}(x)\cdot \vec{I}(x)}{(x-\alpha)^{n_\alpha}} + \vec{Q}^{(\infty)}(x) \cdot \vec{I}(x),
\end{align}
where $n_\alpha$ is the prescribed maximal pole order, $Q_i^{(\alpha)}(x)$ is a polynomial of degree below $n_\alpha$, and the polynomial row $\vec Q^{(\infty)}(x)$ describes the regular contribution to the coefficients. The pole orders and numerator degree bounds are taken as inputs, since they do not depend on the choice of the phase-space point or of the finite-field primes, and hence they can be obtained cheaply.

By (P1), the contribution of $F$ from the analytic region is finite at $\alpha$, which demands the analyticity constraint
\begin{align}\label{eq:finiteness}
\vec{Q}^{(\alpha)}(x)\cdot \vec I(x) = \mathcal{O}\!\left((x-\alpha)^{n_\alpha}\right)
\end{align}
in the neighborhood of $\alpha$. Thus, the analytic contributions of the MIs must combine into a numerator that cancels the prescribed pole. The allowed polynomial numerator rows are closed under addition and multiplication by polynomials, and hence form a \emph{module}~\cite{CoxLittleOShea:2005}. Writing the Gr\"obner basis (GB) of the module as $\vec{v}^{(\alpha)}_{k}(x)$, we express the reduction as
\begin{align}\label{eq:groebner}
F = \sum_{\alpha\in\mathcal{S}_A} \sum_{k=1} Q^{(\alpha)}_{k}(x) \frac{\vec{v}^{(\alpha)}_{k}(x)\cdot\vec{I}(x)}{(x-\alpha)^{n_\alpha}} + \vec{Q}^{(\infty)}(x) \cdot \vec{I}(x),
\end{align}
where the polynomial multipliers $Q^{(\alpha)}_k(x)$ are chosen so that each component of the total numerator $\sum_k Q^{(\alpha)}_k\vec v^{(\alpha)}_k$ has degree below $n_\alpha$.

So far, only the analytic regions at finite $\alpha\in\mathcal{S}_A$ have been used. The leading behavior of $F$ and the MIs at infinity supplies further constraints, restricting $\vec Q^{(\infty)}$ and correlating the finite-pole parameters. Analyticity thus filters the reduction solution space dramatically, leaving far fewer unknown parameters. The parameters then can be fixed by IBP evaluations or asymptotic expansions.

\sect{Generating the numerator module from the differential equation}
\label{sec:gb-from-de}
Since the local analytic structure of the MIs is fixed by the DEs, the module from analyticity is also determined by the DEs and does not depend on particular boundary values. Proofs and further details of this section are given in the Supplemental Material (SM), Secs.~\ref{sec:sm-why-de} and~\ref{sec:sm-numerator-module}.

Repeated differentiation of DEs gives derivative-rows
\begin{align}
 \partial^m_x I_i&=\vec r_i^{[m]}(x)\cdot\vec I(x),\nonumber
  \\ \vec r_i^{[0]}&=\vec e_i,\quad
  \vec r_i^{[m+1]}=\partial_x\vec r_i^{[m]}+\vec r_i^{[m]}A,
  \label{eq:derivative-row-recursion}
\end{align}
where $\vec e_i$ is the unit vector of the component $i$, $\vec r_i^{[m]}(x)$ is rational with poles confined to $\mathcal{S}_A$, and each differentiation raises its pole order at $\alpha$. The rows with $m\le m^{\min}_{i,\alpha}$ then contain the pole structures required to compensate the denominator $(x-\alpha)^{n_\alpha}$, where $m^{\min}_{i,\alpha}$ is the smallest derivative order of $I_i$ supplying the pole order $n_\alpha$ at $\alpha$, for example, $m^{\min}_{i,\alpha}=n_\alpha$ is enough for a Fuchsian pole of $A$ from experience. Since the analytic region of $\partial^m_x I_i$ is $\mathcal O(1)$, we have
\begin{align}
 (x-\alpha)^{n_\alpha}\partial^m_x I_i=\mathcal O\!\left((x-\alpha)^{n_\alpha}\right).
\end{align}
%Because the numerator module can be generated by the DEs, for any allowed polynomial numerator there must exist polynomials $f_{im}^{(\alpha)}(x)$ such that
%\begin{align}\label{eq:reduction-from-derivative-rows}
%   \frac{\vec Q^{(\alpha)}(x)\cdot\vec I(x)}{(x-\alpha)^{n_\alpha}} =
% \sum_{i,\,m\le m^{\min}_{i,\alpha}}f_{im}^{(\alpha)}(x)\,\partial^m_x I_i =\mathcal O(1),
%\end{align}
%where $\mathcal O(1)$ refers to the analytic region. As the MIs are linearly independent over $\mathbb{Q}(x)$, we have
%\begin{align}
% \vec Q^{(\alpha)}(x)
% =(x-\alpha)^{n_\alpha}
%  \sum_{i,\,m\le m^{\min}_{i,\alpha}}f_{im}^{(\alpha)}(x)\vec r_i^{[m]}(x),
% \label{eq:numerator-from-derivative-rows}
%\end{align}
Comparing with Eq.~\eqref{eq:finiteness}, the rows $(x-\alpha)^{n_\alpha}\vec r_i^{[m]}(x)$ with $m\le m^{\min}_{i,\alpha}$ form a generating set of the module. Reducing them gives the GB $\{\vec v^{(\alpha)}_k\}$.

If one prefers to write all coefficients over the complete common denominator $\mathcal Q(x)=\prod_{\alpha\in\mathcal{S}_A}(x-\alpha)^{n_\alpha}$, the analyticity condition \eqref{eq:finiteness} must be imposed at every zero of $\mathcal Q$ simultaneously. The global module is generated by lifting the local generating rows, i.e., replacing the prefactor $(x-\alpha)^{n_\alpha}$ in each row by $\mathcal Q$. Since the factors of $\mathcal Q$ are coprime, the lifted row $\mathcal Q\,\vec r_i^{[m]}$ carries the prescribed pole power at every zero of $\mathcal Q$. It is polynomial except at zeros $\beta$ with $m>m^{\min}_{i,\beta}$, where a residual denominator remains; the global module consists of the polynomial combinations in which all residual denominators cancel. Reducing them gives the GB $\{\vec v^{(\mathcal Q)}_k\}$, and the size of the basis matches the number of MIs.

\sect{Single-variable reduction examples}
\label{sec:single-variable}
Unless stated otherwise, the numerical tests use $\epsilon=1/7$, over the finite field $\mathbb F_{2147483629}$, imposing constraints from the analytic region at finite singularities and from all contributing regions at infinity.\footnote{Non-analytic regions at finite singularities are not used to impose homogeneous constraints on the ansatzes: their systematic inclusion is left to future work, and their use in fixing the unknown parameters is discussed in the section \emph{Determination from asymptotic expansions} below.}

\begin{figure}[htbp]
%\begin{center}
    \includegraphics[width=6cm]{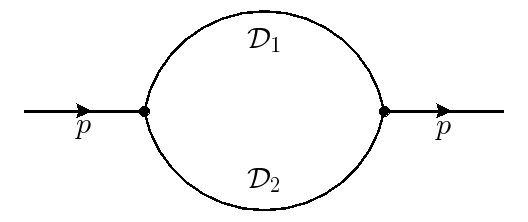}
\caption{\label{fig:bubble}
Feynman diagram of the one-loop two-point integral.}
%\end{center}
\end{figure}
As an elementary example, consider the equal-mass bubble shown in Fig.~\ref{fig:bubble}, with
\begin{align}
 \mathcal D_1=\ell^2-1,\qquad \mathcal D_2=(\ell+p)^2-1,\qquad s=p^2.
  \label{eq:bubble-family}
\end{align}
For the MIs $\vec I=(I_{10},I_{11})$, the DEs read $\partial_s I_{10}=0$ and $\partial_s I_{11}=[-12I_{10}+(14-s)I_{11}]/[7s(s-4)]$. For the two-dot target $I_{22}$, take the prescribed denominator $\mathcal Q=s(s-4)^2$ and the componentwise numerator degree cap $(1,1)$.

The DEs imply the local relations for the analytic region as $-6I_{10}+7I_{11}=\mathcal O(s)$ and $(s-22)I_{10}+15I_{11}=\mathcal O((s-4)^2)$. Applying the derivative-row construction gives the GB
\begin{align}
 \vec v^{(\mathcal Q)}_1=(s(s-4)^2,0),\,
 \vec v^{(\mathcal Q)}_2=(4s^2-25s-90,105),
  \label{eq:bubble-full-generators}
\end{align}
with $\vec N_{\mathcal Q}=f_1\vec v^{(\mathcal Q)}_1+f_2\vec v^{(\mathcal Q)}_2$. The degree cap of $\vec N_{\mathcal Q}$ then requires $f_1=c$, $f_2=a+bs$ with $c+4b=0$ and $-8c+4a-25b=0$ to cancel higher powers of $s$, so the bounded numerator is one-dimensional,
\begin{align}
 I_{22}=\frac{\kappa}{s(s-4)^2}\left[(-21s+30)I_{10}+(20s-35)I_{11}\right].
  \label{eq:bubble-bounded-reduction}
\end{align}
The module constraints thus reduce the four coefficients of two general linear numerators to a single parameter, and one generic IBP point fixes $\kappa=-4/49$.

We extend the tests to higher dots in the equal-mass bubble, unequal-mass sunset, and planar double-box families, covering $2$--$10$, $2$--$8$, and $2$--$6$ dots, respectively. Throughout, the prescribed denominators and numerator degree caps are kept fixed, and we compare the number of unknown parameters before and after imposing the analyticity constraints, together with the number of numerical IBP points needed to determine these parameters.

The compression is substantial in all three families. Every bubble target up to 10 dots retains only a single overall normalization. For the sunset at 8 dots, the ansatz shrinks from 148 parameters to 9, with the required IBP points dropping from 23 to 2. For the double box, at 6 dots the parameter count falls from 189 to 65 and the IBP points from 14 to 4. As will be seen in the multivariable discussion below, these 4 IBP points in fact suffice to reconstruct the full $(s,t)$ dependence. Full results are collected in SM Sec.~\ref{sec:sm-single-variable-details}.

\sect{Bootstrapping the differential equation}
\label{sec:debootstrap}
Thus far, the DE matrix $A(x)$ has been assumed known and used to derive the analyticity constraints. The same bootstrap can also determine $A(x)$ itself.

Ordering MIs by sector, the DE has the block-triangular form
\begin{align}
 \partial_x
  \begin{pmatrix}\vec I_S\\ \vec I_{\rm sub}\end{pmatrix}
  =
  \begin{pmatrix}A_{11}&A_{12}\\ 0&A_{22}\end{pmatrix}
  \begin{pmatrix}\vec I_S\\ \vec I_{\rm sub}\end{pmatrix},
  \label{eq:sector-triangular-de}
\end{align}
where $A_{11}$ acts on the current sector $\vec I_S$, $A_{12}$ and $A_{22}$ on the subsector MIs $\vec I_{\rm sub}$. In a bottom-up construction, $A_{22}$ is already known from the lower sectors, and $A_{11}$ can be obtained independently from a maximal-cut calculation~\cite{Primo:2016ebd,Bosma:2017ens}, where subsector integrals are absent. Hence only the off-diagonal block $A_{12}$ remains to be bootstrapped.

At the pole $(x-\alpha)^{n_\alpha}$, if $\alpha$ is not a pole of $A_{11}$, the subsector numerator module generated from $A_{22}$, with local GB $\{\vec v_{A_{22},l}^{(\alpha)}\}$ (for a non-pole of $A_{22}$, simply $\{(x-\alpha)^{n_\alpha}\vec e_j\}$), already suffices for the ansatz of $A_{12}$. However, if $\alpha$ is a pole of $A_{11}$, the singular contributions of $A_{11}\vec I_S$ and $A_{12}\vec I_{\rm sub}$ must cancel each other, and the subsector module alone is insufficient. Denote $\vec{A}_{11,r}^{(\alpha)}$ as $r$th rows of the numerators associated with the denominator $(x-\alpha)^{n_\alpha}$ of the block $A_{11}$. The cancellation then demands subsector numerators $\vec{u}_{A_{12},r}^{(\alpha)}$, called \emph{extensions}, satisfying 
\begin{align}
 \vec{A}_{11,r}^{(\alpha)}(x)\cdot\vec I_S(x)
 +\vec{u}_{A_{12},r}^{(\alpha)}(x)\cdot\vec I_{\rm sub}(x)
 =\mathcal O\!\left((x-\alpha)^{n_\alpha}\right),
 \label{eq:extension-analyticity}
\end{align}
which is precisely the analyticity condition in Eq.~\eqref{eq:finiteness}.

Because the syzygy relation $\sum_r s_r(x)\vec{A}_{11,r}^{(\alpha)}(x) 
 =\mathcal O\!\left((x-\alpha)^{n_\alpha}\right)$ among the diagonal-block numerators exists, the $\vec{u}_{A_{12},r}^{(\alpha)}$ cannot be chosen arbitrarily. To see this, multiplying Eq.~\eqref{eq:extension-analyticity} by the syzygy coefficients $s_r(x)$ and summing over $r$, the diagonal contribution cancels by the syzygy relation itself, leaving a pure-subsector contribution. Then it must lie in the subsector module:
\begin{align}
 \sum_r s_r(x)\vec{u}_{A_{12},r}^{(\alpha)}(x)
 =\sum_l h_l(x)\vec v_{A_{22},l}^{(\alpha)}(x).
 \label{eq:extension-module-compatibility}
\end{align}
%Conversely, any set of $\vec{u}_{A_{12},r}^{(\alpha)}$ satisfying Eq.~\eqref{eq:extension-module-compatibility} for every syzygy defines a consistent extension $\vec{A}_{11,r}^{(\alpha)} \mapsto \vec{u}_{A_{12},r}^{(\alpha)}$: two representations of the same diagonal source differ by a syzygy, whose image is absorbed by the subsector module. Hence Eq.~\eqref{eq:extension-module-compatibility} is necessary and sufficient, and solving it gives the allowed $u_{A_{12}}^{(\alpha)}$. 
Solving the compatibility conditions \eqref{eq:extension-module-compatibility}
for every syzygy relation gives a finite generating set of extensions, represented by a basis
$\vec g^{(\alpha)}_{A_{12},k}$. Every allowed $\vec u^{(\alpha)}_{A_{12},r}$
is a polynomial combination of the basis. The off-diagonal numerator ansatz at $\alpha$ is
then obtained by combining these extensions with the subsector module. 
(See SM Sec.~\ref{sec:sm-extension-compatibility} for the explicit
compatibility algorithm).

For the AB of DEs, the pole structure of each diagonal block determines the extensions of the corresponding off-diagonal blocks, so spurious poles or higher-order poles enlarge the extensions. Therefore, it is preferable to choose the basis sector by sector such that the diagonal blocks are Fuchsian whenever possible~\cite{Lee:2014ioa,Gituliar:2017vzm} and contain as few spurious divisors as possible (see SM Sec.~\ref{sec:sm-basis-details} for a quantitative comparison of MI bases).

\begin{figure}[htbp]
%\begin{center}
    \includegraphics[width=6cm]{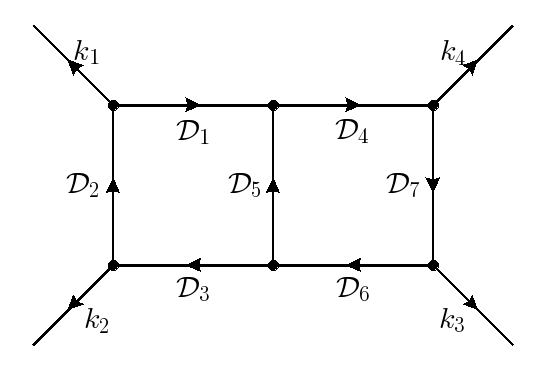}
\caption{\label{fig:doublebox}
Feynman diagram of two-loop double box.}
%\end{center}
\end{figure}

\sect{Bootstrapping differential equations examples}
\label{sec:de-bootstrap-examples}
We first bootstrap the auxiliary-mass DE of the massless planar double box of Fig.~\ref{fig:doublebox} (see SM Sec.~\ref{sec:sm-doublebox-details} for details of the definition). The auxiliary mass is inserted through $\mathcal D_1\to\mathcal D_1-\eta$. As shown in Table~\ref{tab:de-bootstrap-summary}, the AB method reduces the ansatz from 212 to 73 parameters and the IBP points requirement from 8 to 2. Here, the dimension change counts the ansatz parameters before and after imposing the numerator-module constraints and extension compatibility, and the IBP points change the numerical IBP points needed to fix the unknown parameters. The top-sector off-diagonal block is usually among the most difficult parts of the DE reconstruction. The double-box top sector contains two MIs, and the corresponding two rows of $A_{12}$ have trivial ansatzes with 30 and 80 parameters, reduced by AB to 10 and 12.

\begin{table}[t]
\caption{Ansatz dimensions and shared-point costs for mass-dependent DE bootstrap. The double-box (25 MIs) covers all non-source off-diagonal blocks and the double-pentagon (379 MIs) covers only the top-sector off-diagonal block.}
\label{tab:de-bootstrap-summary}
\begin{ruledtabular}
\begin{tabular}{lcc}
family&dimension change&IBP points change\\
double box&$212\to73$&$8\to2$\\
double pentagon&$168{,}605\to10{,}615$&$182\to5$\\
\end{tabular}
\end{ruledtabular}
\end{table}

\begin{figure}[htbp]
    \includegraphics[width=6cm]{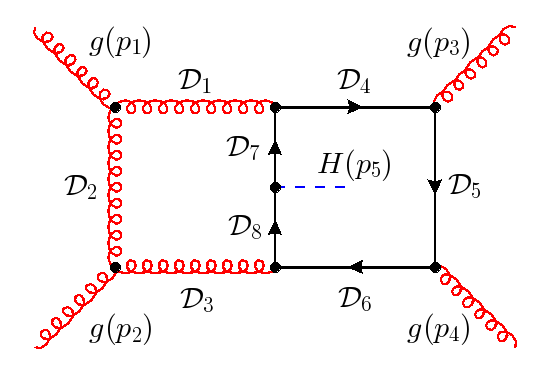}
\caption{\label{fig:hjj_topology}
Topology of the nonplanar double-pentagon family.}
\end{figure}

We also test a nonplanar double-pentagon family contributing to $g(p_1)+g(p_2)\to g(p_3)+g(p_4)+H(p_5)$ through a massive top-quark loop, shown in Fig.~\ref{fig:hjj_topology}, with $m_t^2$ as the differential variable. The family contains 379 MIs, with 14 in the top-sector and 365 in its subsectors. We reconstruct the $14\times365$ off-diagonal block $A_{12}$. As shown in Table~\ref{tab:de-bootstrap-summary}, assuming known denominators, the AB method significantly reduces the ansatz from 168,605 to 10,615 parameters and the IBP points requirement from 182 to 5.

The same example permits a direct comparison with the conventional auxiliary-mass-flow construction, summarized in Table~\ref{tab:hjj-amflow-timing}: interpreting $m_t^2$ as the auxiliary mass of {\tt AMFlow} reproduces the family above. With {\tt Blade} as the IBP reducer, the measured construction required 372 IBP points per prime over 109 finite-field primes. Incorporating AB, only 5 points are needed per prime. This reduces the construction cost to $5/372$ of its original value, from $52.2$ to approximately $0.7\,\mathrm{CPU\,h}$, below the cost of the numerical solution itself.

\begin{table}[t]
  \caption{
    Comparison for the 379-MI double-pentagon family at fixed $\epsilon=1/3$. Both rows
    use the same 109 finite-field primes.  Times are in CPU hours.
  }
  \label{tab:hjj-amflow-timing}
\begin{ruledtabular}
  \begin{tabular}{lccc}
    Method
      & IBP points
      & $t_{\text{SetDE}}$
      & $t_{\text{SolveDE}}$\\
    {\tt Blade}+{\tt AMFlow}
      & 372 & 52.2 & 0.9 \\
    AB+{\tt Blade}+{\tt AMFlow}
      & 5 & 0.7 & 0.9 \\
  \end{tabular}
  \end{ruledtabular}
\end{table}

\sect{Determination from asymptotic expansions}
\label{sec:regional-matching} 
Asymptotic expansions based on the expansion-by-regions method provide an alternative to IBP sampling for reduction~\cite{Beneke:1997zp,Pak:2010pt,Jantzen:2011nz,Liu:2018dmc}.
Each expansion order of a region yields a few independent equations, so an ansatz with many unknown parameters needs correspondingly many orders. The complexity of the expansions grows rapidly with the order, which makes the reconstruction of a trivial ansatz costly. The compact AB ansatz, in contrast, has few enough parameters that very low-order expansions suffice, which are much cheaper to compute and may offer a more efficient route than numerical IBP sampling.

We illustrate this on the double-box top sector, continuing the example above. The two rows of the off-diagonal block $A_{12}(\eta)$ contain 10 and 12 unknown parameters, respectively. Expanding the regions at $\eta=0,1,\infty$ to leading order already provides enough independent equations to fix all of them, so the full block is determined from the leading regional expansions alone, which are much easier to obtain than the IBP reduction of the original family. The details are given in SM Sec.~\ref{sec:sm-regional-details}.

\sect{Multivariable generalization}
\label{sec:multivariable}
For the multivariable case, the one-variable derivative label $[m]$ introduced above is replaced by a multi-index $[\boldsymbol\beta]$, which includes the mixed derivative rows generated for the numerator module construction. Differentiation paths are identified and checked by the flatness conditions
\begin{align}
 \partial_{x}A_y-\partial_{y}A_x+[A_y,A_x]=0.
  \label{eq:multivariable-flatness}
\end{align}
The remaining module and degree-cap steps of AB are unchanged. The infinity constraints, in contrast, require new ingredients: the Newton-fan strata and generic endpoint regions, following the geometric approach to expansion by regions~\cite{Pak:2010pt,Jantzen:2011nz}.

We test the multivariable generalization on the unequal-mass bubble, on the unequal-mass sunset with kinematic variable $s$ and one auxiliary mass $\eta$, and on the one-external-mass double box with kinematic variables $(s,t)$. The tests cover the same targets as the single-variable tests above.

For every tested bubble and sunset target, only one overall normalization remains: at 8 dots, the sunset parameter count drops from 2406 to 1 and the required IBP points from 378 to 1. For the double box at 6 dots, the parameter count decreases from 1605 to 62 and the required IBP points from 120 to 4, almost as compact as the single-variable ansatz at the same target. In relative terms, the constraints remove approximately $96$--$97\%$ of the parameters for the same targets, compared with $66$--$70\%$ on the $t=3$ slice. The reason is that the mixed derivative rows now impose the compatibility conditions carried by the $t$ dependence, which a single-variable construction misses. Full results are given in SM Sec.~\ref{sec:sm-multivariable-details}.

%%%%%%%%%%%%%%%%%%%%%%%%%%%%%%%%%%%%%%%%%%%%%%%%%

\sect{Summary and outlook}
We have developed the analyticity bootstrap, which uses the analyticity of Feynman integrals, encoded in the solution space of their differential equations (DEs), to constrain both the integral reduction and the DEs themselves. The constraints take an explicit algebraic module form: the Gr\"obner basis is generated from the DEs through derivative rows, in one or several variables, and the off-diagonal DE blocks are constrained along with the sector extensions.

We tested the method on several families, both for reduction and for constructing DEs. The residual ansatz is so compact that a few IBP points determine it completely: on the nonplanar double-pentagon of Higgs-dijets production, the reconstruction of the DEs needs only 5 IBP points per prime instead of 372, and on the two-variable double box only 4 instead of 120. The compact ansatz can likewise be determined from low-order asymptotic expansions, without any IBP reduction of the original family. 

The method applies not only to the reduction of individual Feynman integrals: the reduction of Feynman amplitudes can be carried out in exactly the same way, opening the prospect of complete amplitude results from the analyticity bootstrap. It is thus expected to provide essential support for solving state-of-the-art problems in colliders and gravitational waves.

%%%%%%%%%%%%%%%%%%%%%%%%%%%%%%%%%%%%%%%%%%%%%%%%%
\sect{Acknowledgments}
We thank B. Feng, W.-H. Wu and Y. Zhang for useful discussion. The work is in part supported by the National Natural Science Foundation of China (No.~12325503) and the High-performance Computing Platform of Peking University. X.~L. is also supported by the China Postdoctoral Science Foundation (No.~8206301723).

%%%%%%%%%%%%%%%%%%%%%%%%%%%%%%%%%%%%%%%%%%%%%%%%%
%%%%%%%%%%%%%%%%%%%%%%%%%%%%%%%%%%%%%%%%%%%%%%%%%
%\bibliographystyle{apsrev4-2}
%\bibliography{references}
%apsrev4-2.bst 2019-01-14 (MD) hand-edited version of apsrev4-1.bst
%Control: key (0)
%Control: author (72) initials jnrlst
%Control: editor formatted (1) identically to author
%Control: production of article title (-1) disabled
%Control: page (0) single
%Control: year (1) truncated
%Control: production of eprint (0) enabled
%

\newpage

\begin{center}
{\large\bfseries Supplemental Material}
\end{center}

\section{Analyticity constraints from the differential equation}
\label{sec:sm-why-de}

In this section, we detail how the DEs determine the analyticity constraints in Eq.~\eqref{eq:finiteness} of the main text. We impose structural regularity on the full local solution space of the DEs in the analytic region, without specifying numerical boundary values.

To see the statement directly, consider $z=x-\alpha$ and the contribution from the analytic region. At each order, the expansion coefficients can be reduced to a minimal vector of boundary MIs $\vec b=(b_1,\ldots,b_{N_b})^T$, giving
\begin{align}
 \vec I_{\mathrm{an}}(z)=\sum_{r=0}^{\infty}z^r \Theta_r\vec b,
  \label{eq:boundary-master-expansion}
\end{align}
where the $N\times N_b$ matrix $\Theta_r$ gives the coefficients of the boundary masters in the $r$th-order expansion, which are determined recursively by the DEs~\cite{Lee:2017qql,Hidding:2020ytt}. The numerator in Eq.~\eqref{eq:finiteness} has degree below $n_\alpha$, so write
\begin{align}
 \vec{Q}^{(\alpha)}(z)=\sum_{j=0}^{n_\alpha-1}z^j\vec q_j,
  \label{eq:local-numerator-coefficients}
\end{align}
Multiplication gives
\begin{align}
 \vec{Q}^{(\alpha)}(z)\cdot\vec I_{\mathrm{an}}(z)=\sum_{r=0}^{\infty}z^r\left[\sum_{j=0}^{\min(r,n_\alpha-1)}\vec q_j \Theta_{r-j}\right]\vec b.
  \label{eq:local-numerator-convolution}
\end{align}
Equation~\eqref{eq:finiteness} requires the bracket in Eq.~\eqref{eq:local-numerator-convolution} to vanish for $0\leq r<n_\alpha$. Since the boundary MIs $\vec b$ are linearly independent, the vanishing orders give
\begin{align}
 \sum_{j=0}^{\min(r,n_\alpha-1)} \vec q_j \Theta_{r-j} = 0,\quad 0\leq r<n_\alpha,
\end{align}
which constrain the numerator coefficients $\vec q_j$. Therefore, the analytic constraints that $\vec{Q}^{(\alpha)}(x)$ must satisfy are generated by the DEs.

\section{Numerator modules over a complete denominator}
\label{sec:sm-numerator-module}

\subsection{Module and Gr\"obner-basis definitions}
\label{sec:sm-module-definition}

For standard background on polynomial modules, Gr\"obner bases, and syzygies, see Ref.~\cite{CoxLittleOShea:2005}.

Let $\vec I=(I_1,\ldots,I_N)^{\mathsf T}$ be a column of $N$ master integrals, and let $\boldsymbol x=(x_1,\ldots,x_n)$ denote the $n$ active variables. The superscript $\mathsf T$ denotes transpose. The coefficient field $\mathbb F$ contains the dimensional regulator $\epsilon$ and all parameters held fixed under differentiation. We use the polynomial ring $R$ and its fraction field $K$, the field of rational functions in $\boldsymbol x$:

\begin{equation}
\begin{gathered}
R=\mathbb F[x_1,\ldots,x_n],\qquad K=\operatorname{Frac}(R),
 \\
 \mathcal Q=\prod_{a=1}^{\sigma}q_a^{n_a}\in R\setminus\{0\}.
 \label{eq:sm-ring-denominator}
\end{gathered}
\end{equation}

Here $\mathcal Q$ is the complete denominator, with $\sigma$ distinct irreducible polynomial factors $q_a$ and positive integer exponents $n_a$. The notation $R^{1\times N}$ denotes polynomial rows with $N$ components; $\vec e_i$ is its unit row in component $i$. An $R$-module of such rows is a set closed under addition and multiplication by polynomials in $R$. Thus module generators may be multiplied by polynomials, whereas a vector-space basis over $\mathbb F$ permits only coefficients in $\mathbb F$.

At a generic smooth point of the divisor $q_a=0$, use $q_a$ as a local transverse coordinate. Let $\mathcal O_a$ denote the ring of formal power series with nonnegative powers of this coordinate, with tangential parameters in the coefficient field. These series describe regular local germs, i.e., local expansions with no transverse pole. Let $\mathscr H_a\subset\mathcal O_a^N$ be the column-vector solution germs of the DE in the analytic region. The numerator module is

\begin{equation}
\begin{gathered}
\mathcal N_{\mathcal Q}
 =\left\{\vec N\in R^{1\times N}\;\middle|\;
 \begin{aligned}
 &\vec N\cdot\vec h\in q_a^{n_a}\mathcal O_a\\
 &\text{for every }a\text{ and }\vec h\in\mathscr H_a
 \end{aligned}\right\}.
 \label{eq:sm-analytic-numerator-module}
\end{gathered}
\end{equation}

The dot denotes row--column contraction. Membership in $q_a^{n_a}\mathcal O_a$ means divisibility by $q_a^{n_a}$ in the local power series. Since $\mathcal Q/q_a^{n_a}$ is invertible there, the condition requires the first $n_a$ transverse Taylor coefficients of $\vec N\cdot\vec h$ to vanish, making $(\vec N/\mathcal Q)\cdot\vec h$ regular. It defines an $R$-submodule containing $\mathcal Q R^{1\times N}$. Polynomial rings over fields are Noetherian, so this submodule has a finite generating set. No numerator degree bound is imposed yet.

To construct numerators from derivatives, write $\partial^{\boldsymbol\beta}=\partial_{x_1}^{\beta_1}\cdots\partial_{x_n}^{\beta_n}$, where $\boldsymbol\beta=(\beta_1,\ldots,\beta_n)$ is a tuple of nonnegative integers. Its total derivative order is $|\boldsymbol\beta|=\sum_a\beta_a$. Reduction to the master basis gives rational rows

\begin{equation}
\begin{gathered}
\partial^{\boldsymbol\beta}I_i
 =\vec r_i^{[\boldsymbol\beta]}\cdot\vec I,
 \qquad \vec r_i^{[\boldsymbol\beta]}\in K^{1\times N},
 \qquad \vec r_i^{[\boldsymbol0]}=\vec e_i.
 \label{eq:sm-derivative-rows}
\end{gathered}
\end{equation}

Let $\Omega$ be the finite list of pairs $(i,\boldsymbol\beta)$ selected as sources, including all zeroth-order rows. Their polynomial span and its polynomial output after multiplication by $\mathcal Q$ are

\begin{equation}
\begin{gathered}
\mathscr D_\Omega
 =\sum_{(i,\boldsymbol\beta)\in\Omega}R\,\vec r_i^{[\boldsymbol\beta]},
 \\
 \mathcal M_{\mathcal Q,\Omega}
 =(\mathcal Q\mathscr D_\Omega)\cap R^{1\times N}.
 \label{eq:sm-fixed-denominator-definition}
\end{gathered}
\end{equation}

The sum denotes all polynomial linear combinations of the source rows, and $\cap$ retains the combinations whose every component is polynomial. Individual summands may have residual denominators after multiplication by $\mathcal Q$; these denominators need only cancel in the complete sum.

Analyticity is preserved by differentiation, so $\mathcal M_{\mathcal Q,\Omega}\subseteq\mathcal N_{\mathcal Q}$. The source set is called complete for $\mathcal Q$ when equality holds. The construction below computes $\mathcal M_{\mathcal Q,\Omega}$ exactly for the supplied sources; with a complete source set, it gives the analyticity module $\mathcal N_{\mathcal Q}$.

A module monomial is $\boldsymbol x^{\boldsymbol\gamma}\vec e_i$, where $\boldsymbol x^{\boldsymbol\gamma}=\prod_a x_a^{\gamma_a}$ and $\gamma_a$ are nonnegative integers. Fix an order on these monomials that is compatible with multiplication and has no infinite descending chain. The leading term $\operatorname{LT}(\vec v)$ is the largest nonzero term of a row $\vec v$, including its coefficient and component position. A finite set $G\subset\mathcal M_{\mathcal Q,\Omega}$ is a module Gr\"obner basis when

\begin{equation}
\begin{gathered}
\operatorname{LT}(\mathcal M_{\mathcal Q,\Omega})
 =\langle\operatorname{LT}(\vec v):\vec v\in G\rangle_R,
 \\
 \vec N\in\mathcal M_{\mathcal Q,\Omega}
 \quad\Longleftrightarrow\quad \operatorname{NF}_{G}(\vec N)=0.
 \label{eq:sm-gb-definition}
\end{gathered}
\end{equation}

The brackets $\langle\cdots\rangle_R$ mean polynomial span, and the left-hand side is the module generated by leading terms of all nonzero rows in $\mathcal M_{\mathcal Q,\Omega}$. The normal form $\operatorname{NF}_{G}$ is obtained by subtracting polynomial multiples of basis rows until no remaining monomial is divisible by a leading monomial of $G$ in the same component. For a fixed Gr\"obner basis this gives the unique remainder in the standard monomials. The resulting $\mathbb F$-linear remainder map tests membership and will impose the finite bounds in Sec.~\ref{sec:sm-bounded-intersection}.

\subsection{Freeness in one variable}
\label{sec:sm-freeness}

In one variable, the unrestricted global numerator module is a free module of rank $N$. To prove this, let $R=\mathbb F[x]$, where $\mathbb F$ contains $\epsilon$ and all fixed parameters, and denote by $\mathcal N_{\mathcal Q}$ the polynomial numerator module satisfying the finite-point analyticity conditions. Every row $\mathcal Q\vec e_i$ belongs to this module: after division by $\mathcal Q$, it selects $I_i$, whose contribution from the analytic region is regular. Thus
\begin{align}
 \mathcal Q R^{1\times N}\subseteq\mathcal N_{\mathcal Q}
  \subseteq R^{1\times N}.
  \label{eq:global-module-inclusions}
\end{align}
Since $R$ is a principal ideal domain, every submodule of the free module $R^{1\times N}$ is free, with rank at most $N$. The $N$ rows $\mathcal Q\vec e_i$ are linearly independent over $R$ because $\mathcal Q\ne0$, so the rank is also at least $N$. Hence $\mathcal N_{\mathcal Q}$ is free of rank $N$ and admits a polynomial basis with exactly $N$ elements.

\subsection{Gr\"obner construction by expanded pole states}
\label{sec:sm-pole-state}

The calculation uses two polynomial-module eliminations. The first identifies different formal states representing the same rational row; the second extracts the polynomial combinations of the derivative sources. Both use finite generating matrices over $R$.

\emph{1. Enumerate the residual pole states.} Extend the list of factors to $q_1,\ldots,q_\tau$ so that it includes every denominator factor of the supplied derivative rows; factors absent from $\mathcal Q$ have exponent zero in $\mathcal Q$. Here $\tau\geq\sigma$ is the length of this extended list. For the $j$th component $r_{ij}^{[\boldsymbol\beta]}$ of a source row, cancel common factors to write

\begin{equation}
\begin{gathered}
\mathcal Q r_{ij}^{[\boldsymbol\beta]}
 =\frac{p_{ij\boldsymbol\beta}}
        {\prod_{a=1}^{\tau}q_a^{\rho_{ij\boldsymbol\beta,a}}},
 \qquad p_{ij\boldsymbol\beta}\in R.
 \label{eq:sm-residual-component}
\end{gathered}
\end{equation}

The tuple $\boldsymbol\rho_{ij\boldsymbol\beta}$ records the nonnegative residual pole exponents. A polynomial component has tuple $\boldsymbol0$; for a zero component take both its numerator and tuple to be zero. For each master component $j$, form a finite set $\mathcal R_j$ containing $\boldsymbol0$ and every tuple $\boldsymbol\rho$ with $0\leq\rho_a\leq\rho_{ij\boldsymbol\beta,a}$ for some source $(i,\boldsymbol\beta)$. This is the componentwise downward closure of the observed tuples.

Introduce one independent formal basis symbol $e_{j,\boldsymbol\rho}$ for each listed state. The free module $E$ consists of polynomial linear combinations of these symbols, and its base submodule $E_0$ contains only zero-pole states:

\begin{equation}
\begin{gathered}
E=\bigoplus_{j=1}^{N}\ \bigoplus_{\boldsymbol\rho\in\mathcal R_j}
       R e_{j,\boldsymbol\rho},
 \qquad E_0=\bigoplus_{j=1}^{N}R e_{j,\boldsymbol0},
 \\
 \operatorname{ev}:E\longrightarrow K^{1\times N},\qquad
 e_{j,\boldsymbol\rho}\longmapsto
       \frac{\vec e_j}{\prod_{a=1}^{\tau}q_a^{\rho_a}}.
 \label{eq:sm-expanded-module}
\end{gathered}
\end{equation}

The direct-sum symbol $\bigoplus$ keeps the state positions independent. The map $\operatorname{ev}$ is extended by polynomial linearity: it replaces each formal symbol by its rational row, without evaluating the kinematic variables at numerical points. On $E_0$ it identifies $e_{j,\boldsymbol0}$ with $\vec e_j$. States with $\boldsymbol\rho\neq\boldsymbol0$ are called auxiliary states.

\emph{2. Compute all state relations.} The kernel $J=\ker(\operatorname{ev})$ is the set of state combinations representing the zero rational row. Enumerate $\mathcal R_j$ as $\boldsymbol\rho_{j1},\ldots,\boldsymbol\rho_{jm_j}$, where $m_j=|\mathcal R_j|$ is the number of states in component $j$, and set

\begin{equation}
\begin{gathered}
\lambda_j=\operatorname{lcm}_{\boldsymbol\rho\in\mathcal R_j}
           \prod_{a=1}^{\tau}q_a^{\rho_a},
  \qquad h_{jb}=\frac{\lambda_j}{\prod_{a=1}^{\tau}q_a^{\rho_{jb,a}}},
 \\
 J_j=\left\{\boldsymbol\zeta\in R^{1\times m_j}:
                       \sum_{b=1}^{m_j}\zeta_b h_{jb}=0\right\},
 \\
 J\simeq\bigoplus_{j=1}^{N}J_j.
 \label{eq:sm-scalar-state-kernel}
\end{gathered}
\end{equation}

Here $\operatorname{lcm}$ is a least common multiple of the state denominators; its overall nonzero constant is immaterial. Each $h_{jb}$ is a polynomial. The row $\boldsymbol\zeta=(\zeta_1,\ldots,\zeta_{m_j})$ is a scalar syzygy, meaning a polynomial relation among these polynomials. Embed it into $E$ as $\sum_b \zeta_b e_{j,\boldsymbol\rho_{jb}}$. Multiplication by the nonzero $\lambda_j$ neither creates nor removes a zero rational combination, which proves the kernel formula. The symbol $\simeq$ denotes this identification of modules.

A concrete way to compute $J_j$ requires no bound on the degrees of the unknown $\zeta_b$. Write $\boldsymbol h_j=(h_{j1},\ldots,h_{jm_j})$, let $\vec e_b$ be the $b$th unit row of $R^{1\times m_j}$, and form the graph module

\begin{equation}
\begin{gathered}
\Gamma_j=\left\langle(h_{jb},\vec e_b):
                     1\leq b\leq m_j\right\rangle_R
           \subset R\oplus R^{1\times m_j},
 \\
 \Psi_j=\bigl(\,\boldsymbol h_j^{\mathsf T}\mid I_{m_j}\,\bigr).
 \label{eq:sm-state-graph-matrix}
\end{gathered}
\end{equation}

The rows of the $m_j\times(m_j+1)$ matrix $\Psi_j$ generate $\Gamma_j$; $I_{m_j}$ is the identity matrix. Any row of this module has the form $(\sum_b \zeta_bh_{jb},\boldsymbol\zeta)$. Compute its module Gr\"obner basis with every monomial in the first coordinate ordered above every monomial in the remaining coordinates. Retain precisely the basis rows whose first coordinate is zero, and delete that zero coordinate. The resulting rows generate $J_j$, by elimination. Repeating this for all $j$ and embedding the results gives generators of $J$.

For example, the relations $q_a e_{j,\boldsymbol\rho}-e_{j,\boldsymbol\rho-\boldsymbol\delta_a}$ belong to $J$ whenever $\rho_a>0$, where $\boldsymbol\delta_a$ is the unit tuple in the $a$th pole exponent. They express cancellation of one denominator factor. The graph calculation includes these relations and any additional syzygies; restricting the input to these nearest-state relations alone need not give the full kernel.

\emph{3. Insert the derivative sources.} Lift each complete source row to $E$ by placing its numerator in the corresponding state position:

\begin{equation}
\begin{gathered}
g_{i\boldsymbol\beta}
 =\sum_{j=1}^{N}p_{ij\boldsymbol\beta}
       e_{j,\boldsymbol\rho_{ij\boldsymbol\beta}},
 \\
 L=J+\sum_{(i,\boldsymbol\beta)\in\Omega}R g_{i\boldsymbol\beta}.
 \label{eq:sm-source-lifts}
\end{gathered}
\end{equation}

The generating matrix of $L$ consists of the embedded generators of $J$ followed by the rows $g_{i\boldsymbol\beta}$. Each source has one common polynomial multiplier across all master components, preserving the correlations in the DE. The zeroth-order sources contribute $\mathcal Q e_{j,\boldsymbol0}$. By construction, the image $\operatorname{ev}(L)$, i.e., the set of represented rational rows, equals $\mathcal Q\mathscr D_\Omega$.

\emph{4. Eliminate the auxiliary states.} Order all auxiliary positions above all base positions, irrespective of polynomial degree, and use the chosen order of Sec.~\ref{sec:sm-module-definition} within $E_0$. Compute a module Gr\"obner basis $\widetilde G$ from the generating matrix of $L$. Both this step and the graph calculation can use ordinary module Buchberger reduction: pair rows with the same leading position, multiply them to the least common multiple of their leading monomials, cancel the leading terms, and reduce the resulting row. Append each nonzero remainder and continue until all such pairs reduce to zero.

\begin{equation}
\begin{gathered}
L\cap E_0\simeq\mathcal M_{\mathcal Q,\Omega},
 \qquad G=\widetilde G\cap E_0.
 \label{eq:sm-eliminated-intersection}
\end{gathered}
\end{equation}

Thus $G$ is obtained by retaining exactly the rows of $\widetilde G$ with zero auxiliary components and identifying their base coordinates with polynomial numerator rows. To see why this is the required intersection, let a lift in $L$ represent a polynomial row. Its difference from the corresponding base-state row lies in $J\subset L$, so the base-state row also belongs to $L$. Conversely, every row of $L\cap E_0$ represents a polynomial combination of the scaled sources. The elimination theorem makes the retained rows a Gr\"obner basis of this intersection. Simply deleting nonzero auxiliary entries would not enforce pole cancellation. One may finally reduce $G$ within $R^{1\times N}$.

\subsection{Componentwise denominators and bounded intersections}
\label{sec:sm-bounded-intersection}

For each coefficient $c_i=p_i/d_i$ allowed to be nonzero, prescribe a polynomial denominator $d_i$ dividing $\mathcal Q$ and a finite set $\mathcal X_i$ of allowed exponent tuples for $p_i$. Write $\boldsymbol d=(d_i)$ and $\mathcal X=(\mathcal X_i)$ for these componentwise choices. Their ambient numerator space is

\begin{equation}
\begin{gathered}
R_{\mathcal X_i}
 =\operatorname{span}_{\mathbb F}
   \{\boldsymbol x^{\boldsymbol\gamma}:\boldsymbol\gamma\in\mathcal X_i\},
 \\
 V_{\mathcal Q,\boldsymbol d,\mathcal X}
 =\bigoplus_i\frac{\mathcal Q}{d_i}R_{\mathcal X_i}\vec e_i.
 \label{eq:sm-componentwise-ambient}
\end{gathered}
\end{equation}

Here $\operatorname{span}_{\mathbb F}$ allows arbitrary linear combinations with coefficients in $\mathbb F$. Thus $\mathcal Q/d_i$ divides the complete numerator component $N_i$, while the support or degree cap applies to $p_i=N_i/(\mathcal Q/d_i)$. For example, $\mathcal X_i=\{\boldsymbol\gamma:|\boldsymbol\gamma|\leq\theta_i\}$ imposes the nonnegative integer total-degree cap $\theta_i$, where $|\boldsymbol\gamma|=\sum_a\gamma_a$. The prescribed $d_i$ is an allowed denominator; cancellation may make the actual denominator smaller. Components known to vanish are omitted.

Enumerate the rows $(\mathcal Q/d_i)\boldsymbol x^{\boldsymbol\gamma}\vec e_i$, with $\boldsymbol\gamma\in\mathcal X_i$, as a basis $\vec m_1,\ldots,\vec m_{n_{\mathcal X}}$, where $n_{\mathcal X}=\sum_i|\mathcal X_i|$. A numerator $\sum_{h=1}^{n_{\mathcal X}}\xi_h\vec m_h$ with $\xi_h\in\mathbb F$ already obeys all denominator and support restrictions. The remaining condition is

\begin{equation}
\begin{gathered}
\mathcal B_{\mathcal Q,\boldsymbol d,\mathcal X}
 =\mathcal M_{\mathcal Q,\Omega}\cap V_{\mathcal Q,\boldsymbol d,\mathcal X},
 \\
 \sum_{h=1}^{n_{\mathcal X}}\xi_h\operatorname{NF}_{G}(\vec m_h)=0.
 \label{eq:sm-bounded-normal-forms}
\end{gathered}
\end{equation}

To turn this into a matrix, list all distinct module monomials occurring in the remainders as $\vec\chi_1,\ldots,\vec\chi_Z$, where $Z$ is their number. Define a matrix $F_{\rm rem}\in\mathbb F^{Z\times n_{\mathcal X}}$ by

\begin{equation}
\begin{gathered}
\operatorname{NF}_{G}(\vec m_h)
   =\sum_{r=1}^{Z}(F_{\rm rem})_{rh}\vec\chi_r,
  \qquad F_{\rm rem}\boldsymbol\xi=0.
 \label{eq:sm-remainder-matrix}
\end{gathered}
\end{equation}

The vector $\boldsymbol\xi=(\xi_1,\ldots,\xi_{n_{\mathcal X}})^{\mathsf T}$ contains the unknown ambient coefficients. The kernel of $F_{\rm rem}$ is an ordinary finite-dimensional nullspace over $\mathbb F$. Mapping its basis vectors by $\boldsymbol\xi\mapsto\sum_h\xi_h\vec m_h$ gives independent numerator rows $\vec w_1,\ldots,\vec w_M$. This map is one-to-one because the $\vec m_h$ are independent. If every remainder is zero, the matrix has no rows and the whole ambient space is retained. The rational-coefficient ansatz is

\begin{equation}
\begin{gathered}
\vec c(\boldsymbol x)
 =\frac{1}{\mathcal Q(\boldsymbol x)}
   \sum_{b=1}^{M}\kappa_b\vec w_b(\boldsymbol x),
 \\
 M=\dim_{\mathbb F}\mathcal B_{\mathcal Q,\boldsymbol d,\mathcal X},
 \qquad \kappa_b\in\mathbb F.
 \label{eq:sm-bounded-ansatz}
\end{gathered}
\end{equation}

Here $M$ is the number of unknown parameters and the $\kappa_b$ are fixed by reconstruction. The construction bounds the complete numerator combination, allowing cancellations between module generators, and needs no guessed degree bound on their polynomial multipliers.

\section{Sector extensions over a common denominator}
\label{sec:sm-sector-extension}

\subsection{Diagonal source and subsector quotient}
\label{sec:sm-extension-source}

For the single-variable DE bootstrap, take $R=\mathbb F[x]$. Let $\vec I_S$ contain the $N_S$ masters of the current sector and $\vec I_{\rm sub}$ the $N_{\rm sub}$ masters in its lower sectors. In the notation of Eq.~\eqref{eq:sector-triangular-de} of the main text,

\begin{equation}
\begin{gathered}
\partial_x\vec I_S=A_{11}\vec I_S+A_{12}\vec I_{\rm sub},
 \qquad \partial_x\vec I_{\rm sub}=A_{22}\vec I_{\rm sub}.
 \label{eq:sm-sector-block-definition}
\end{gathered}
\end{equation}

The known rational matrices $A_{11}$ and $A_{22}$ have sizes $N_S\times N_S$ and $N_{\rm sub}\times N_{\rm sub}$; the unknown off-diagonal block $A_{12}$ has size $N_S\times N_{\rm sub}$. Choose a nonconstant monic common denominator $\mathcal Q$, meaning that its leading coefficient is one, which clears $A_{11}$ and is divisible by every prescribed off-diagonal denominator. Define

\begin{equation}
\begin{gathered}
T=R/(\mathcal Q),\qquad \delta_{\mathcal Q}=\deg\mathcal Q,
 \\
 \vec a_r=\operatorname{rem}_{\mathcal Q}(\mathcal Q A_{11,r*}),
 \qquad P_{A_{11}}=\langle\overline{\vec a}_r\rangle_T
       \subset T^{1\times N_S}.
 \label{eq:sm-common-diagonal-source}
\end{gathered}
\end{equation}

The quotient ring $T$ identifies polynomials whose difference is divisible by $\mathcal Q$; addition and multiplication are followed by reduction modulo $\mathcal Q$. Each class has a unique representative of degree below $\delta_{\mathcal Q}$, so $1,x,\ldots,x^{\delta_{\mathcal Q}-1}$ form an $\mathbb F$-basis. A bar denotes the class, and $\operatorname{rem}_{\mathcal Q}$ denotes its polynomial representative, taken componentwise for rows. The symbol $A_{11,r*}$ means the $r$th row of $A_{11}$, with $1\leq r\leq N_S$.

The source module $P_{A_{11}}$ is the $T$-span of the diagonal pole numerators. Its polynomial preimage is $\langle\vec a_r,\mathcal Q\vec e_i\rangle_R$, where the $\vec e_i$ now have $N_S$ components; a Gr\"obner basis of this preimage gives alternative source generators after reduction modulo $\mathcal Q$. Keeping the factors in $\mathcal Q A_{11,r*}$ retains the original pole orders even when $\mathcal Q$ has higher powers. Taking the remainder removes only polynomial coefficient rows after division by $\mathcal Q$.

Apply Sec.~\ref{sec:sm-numerator-module} to the known subsector DE $A_{22}$, obtaining its numerator module $\mathcal N_{A_{22},\mathcal Q}$. Since it contains $\mathcal Q R^{1\times N_{\rm sub}}$, define the finite $T$-modules

\begin{equation}
\begin{gathered}
\overline{\mathcal N}_{A_{22},\mathcal Q}
 =\mathcal N_{A_{22},\mathcal Q}/\mathcal Q R^{1\times N_{\rm sub}},
 \\
 V_{A_{22}}=T^{1\times N_{\rm sub}}/
       \overline{\mathcal N}_{A_{22},\mathcal Q}.
 \label{eq:sm-common-subsector-quotient}
\end{gathered}
\end{equation}

The first quotient is the image of the known subsector numerator module modulo $\mathcal Q$. The second identifies two residue rows when their difference belongs to that image. We write $[\overline{\vec u}]$ for the resulting class of a polynomial row $\vec u\in R^{1\times N_{\rm sub}}$ in $V_{A_{22}}$. Thus two representatives are equivalent precisely when their difference, divided by $\mathcal Q$, gives a regular subsector combination in the analytic region. The quotient retains the classes whose poles still require cancellation with the current sector.

An analytic lift of a source numerator $\vec a_r$ is a polynomial subsector row $\vec u_r$ such that, at every divisor $q_a\mid\mathcal Q$, the paired numerator satisfies

\begin{equation}
\begin{gathered}
\frac{\vec a_r\cdot\vec I_{S,\mathrm{an}}
       +\vec u_r\cdot\vec I_{\mathrm{sub},\mathrm{an}}}{\mathcal Q}
 \in\mathcal O_a.
 \label{eq:sm-common-analytic-lift}
\end{gathered}
\end{equation}

The columns $\vec I_{S,\mathrm{an}}$ and $\vec I_{\mathrm{sub},\mathrm{an}}$ are the current-sector and subsector parts of the analytic solution germs at the divisor under consideration. The condition is imposed on this solution space, as in Sec.~\ref{sec:sm-module-definition}. Two lifts of the same source differ by an element of $\mathcal N_{A_{22},\mathcal Q}$. Consequently, an analytic extension induces a map from $P_{A_{11}}$ to $V_{A_{22}}$ that is linear over $T$, i.e., it respects sums and multiplication by every residue polynomial in $T$.

\subsection{Syzygies and the simultaneous compatibility system}
\label{sec:sm-extension-compatibility}

Choose $g$ generators $\overline{\vec a}_1,\ldots,\overline{\vec a}_g$ of $P_{A_{11}}$ and degree-below-$\delta_{\mathcal Q}$ polynomial representatives $\vec a_r$. They may be the original rows or the alternative generators described above. Their syzygy module records every polynomial relation among these residue rows:

\begin{equation}
\begin{gathered}
\operatorname{Syz}_{\mathcal Q}(P_{A_{11}})
 =\left\{\boldsymbol s\;\middle|\;
       \sum_{r=1}^{g}s_r\vec a_r=0\pmod{\mathcal Q}\right\}.
 \label{eq:sm-common-diagonal-syzygies}
\end{gathered}
\end{equation}

Here $\boldsymbol s=(s_1,\ldots,s_g)\in R^{1\times g}$ is a row of polynomial multipliers, and equality modulo $\mathcal Q$ means that every component of the sum is divisible by $\mathcal Q$. Combining analytic lifts with this relation leaves a regular diagonal contribution. Hence compatibility requires

\begin{equation}
\begin{gathered}
\sum_{r=1}^{g}s_r\vec u_r\in\mathcal N_{A_{22},\mathcal Q}
 \qquad\bigl(\boldsymbol s\in\operatorname{Syz}_{\mathcal Q}(P_{A_{11}})\bigr).
 \label{eq:sm-common-extension-membership}
\end{gathered}
\end{equation}

Conversely, this condition is sufficient for $\overline{\vec a}_r\mapsto[\overline{\vec u}_r]$ to define a $T$-linear map: any two source representations differ by a syzygy and therefore have the same image in $V_{A_{22}}$. The space of compatible maps is

\begin{equation}
\begin{gathered}
\mathcal H_{\mathcal Q}=\operatorname{Hom}_{T}(P_{A_{11}},V_{A_{22}}).
 \label{eq:sm-common-extension-hom}
\end{gathered}
\end{equation}

The notation $\operatorname{Hom}_{T}$ denotes the vector space of all $T$-linear maps $\phi:P_{A_{11}}\to V_{A_{22}}$. These are candidate compatible extensions: the relation condition makes the assignment well defined, while the remaining reconstruction conditions select the physical coefficients.

For an explicit calculation of the syzygies, let $\operatorname{coeff}$ send a polynomial row of degree below $\delta_{\mathcal Q}$ to the column of its coefficients, ordered first by master component and then by powers $1,x,\ldots,x^{\delta_{\mathcal Q}-1}$. Form a matrix $Z_{A_{11}}\in\mathbb F^{\delta_{\mathcal Q}N_S\times\delta_{\mathcal Q}g}$ whose column labeled by $(r,j)$ is

\begin{equation}
\begin{gathered}
 (Z_{A_{11}})_{*,(r,j)}
 =\operatorname{coeff}\!\left(
        \operatorname{rem}_{\mathcal Q}(x^j\vec a_r)\right),
 \\
 1\leq r\leq g,\qquad 0\leq j<\delta_{\mathcal Q}.
 \label{eq:sm-source-relation-matrix}
\end{gathered}
\end{equation}

The asterisk denotes all entries in that column. A nullspace vector with entries $s_r^{(j)}\in\mathbb F$ gives a relation $s_r=\sum_{j=0}^{\delta_{\mathcal Q}-1}s_r^{(j)}x^j$. Map a basis of $\ker Z_{A_{11}}$ into such polynomial rows and adjoin $\mathcal Q\vec e_r$, where $\vec e_r$ now has $g$ components. These rows generate all syzygies over $R$: any multiplier can be divided by $\mathcal Q$, leaving a degree-below-$\delta_{\mathcal Q}$ remainder and a multiple of $\mathcal Q$.

Alternatively, solve for all compatible maps without explicitly exporting the syzygies. Let $d_P=\dim_{\mathbb F}P_{A_{11}}$ and $d_V=\dim_{\mathbb F}V_{A_{22}}$, and choose bases $\omega_1,\ldots,\omega_{d_P}$ and $\bar\omega_1,\ldots,\bar\omega_{d_V}$ of these spaces. The symbols $\omega_i$ and $\bar\omega_j$ denote residue-row classes and quotient classes, respectively. Represent multiplication by $x$ and an unknown $\mathbb F$-linear map $\phi$ by

\begin{equation}
\begin{gathered}
x\omega_i=\sum_{r=1}^{d_P}(X_{A_{11}})_{ir}\omega_r,
  \qquad x\bar\omega_j=\sum_{s=1}^{d_V}(X_{A_{22}})_{js}\bar\omega_s,
  \\
  \phi(\omega_i)=\sum_{j=1}^{d_V}\Phi_{ij}\bar\omega_j.
 \label{eq:sm-common-multiplication-matrices}
\end{gathered}
\end{equation}

The known matrices $X_{A_{11}}$ and $X_{A_{22}}$ have sizes $d_P\times d_P$ and $d_V\times d_V$; the unknown matrix $\Phi$ has size $d_P\times d_V$, with entries in $\mathbb F$. This convention records images in rows. To obtain the source basis, select independent columns from $Z_{A_{11}}$. To obtain a quotient basis, form the coefficient span of $\operatorname{rem}_{\mathcal Q}(x^j\vec v_{A_{22},l})$ for all chosen generators $\vec v_{A_{22},l}$ of $\mathcal N_{A_{22},\mathcal Q}$ and $0\leq j<\delta_{\mathcal Q}$, then extend it to a basis of the ambient residue-row space. The added vectors represent a basis of $V_{A_{22}}$.

For each basis representative, multiply by $x$ and reduce modulo $\mathcal Q$. Re-express the source result in the $\omega_i$ basis to obtain $X_{A_{11}}$. For the quotient, expand in the extended ambient basis and discard the coordinates along $\overline{\mathcal N}_{A_{22},\mathcal Q}$; the remaining coordinates give $X_{A_{22}}$. Equating $\phi(x\omega_i)$ and $x\phi(\omega_i)$ gives

\begin{equation}
\begin{gathered}
X_{A_{11}}\Phi=\Phi X_{A_{22}}.
 \label{eq:sm-common-hom-system}
\end{gathered}
\end{equation}

This is a homogeneous linear system in the $d_Pd_V$ entries of $\Phi$. A basis of its nullspace gives a basis of $\mathcal H_{\mathcal Q}$. Commuting with $x$ enforces commuting with every polynomial in $x$, hence $T$-linearity and all source relations, including annihilation by a polynomial. Since the powers $q_a^{n_a}$ are pairwise coprime in $\mathbb F[x]$, the Chinese remainder theorem identifies $T$ with $\prod_{a=1}^{\sigma}R/(q_a^{n_a})$. One may therefore perform the calculation in separate factor blocks and recombine it without changing the common-denominator construction.

\subsection{Fixed-row extension module and bounded ansatz}
\label{sec:sm-fixed-row-extension}

Let $r_0$ label the particular row of $A_{12}$ being reconstructed, and let $\overline{\vec a}_{r_0}$ be its original diagonal source. This row need not be one of the chosen generators of $P_{A_{11}}$. Expand it in the $\omega_i$ basis and apply each basis map in $\mathcal H_{\mathcal Q}$ to it using the corresponding matrix $\Phi$. Distinct maps may have the same image on this source; keep only linearly independent images. Define

\begin{equation}
\begin{gathered}
\mathcal L_{r_0}
 =\{\phi(\overline{\vec a}_{r_0}):\phi\in\mathcal H_{\mathcal Q}\}
 \subset V_{A_{22}},
 \\
 \mathcal E_{\mathcal Q,r_0}=\pi^{-1}(\mathcal L_{r_0}).
 \label{eq:sm-fixed-row-module}
\end{gathered}
\end{equation}

Here the quotient map $\pi$ and the inverse-image notation mean

\begin{equation}
\begin{gathered}
\pi:R^{1\times N_{\rm sub}}\longrightarrow V_{A_{22}},
 \qquad \vec u\longmapsto[\overline{\vec u}],
 \\
 \pi^{-1}(\mathcal L_{r_0})
 =\{\vec u\in R^{1\times N_{\rm sub}}:
                    \pi(\vec u)\in\mathcal L_{r_0}\}.
 \label{eq:sm-extension-quotient-map}
\end{gathered}
\end{equation}

Thus $\pi$ first reduces modulo $\mathcal Q$ and then removes the pure-subsector numerator class. Its kernel is $\mathcal N_{A_{22},\mathcal Q}$, and $\pi^{-1}$ denotes a set of representatives, not an inverse function. The image space $\mathcal L_{r_0}$ is closed under multiplication by $T$: multiplying any compatible map by an element of $T$ gives another compatible map. Choose polynomial representatives $\vec u_k$ of an $\mathbb F$-basis of this image space. Equivalently, the full extension module decomposes as

\begin{equation}
\begin{gathered}
\mathcal E_{\mathcal Q,r_0}
 =\mathcal N_{A_{22},\mathcal Q}+\sum_k R\vec u_k.
 \label{eq:sm-common-extension-generators}
\end{gathered}
\end{equation}

A Gr\"obner reduction of the pure-subsector generators together with the $\vec u_k$ gives a basis of $\mathcal E_{\mathcal Q,r_0}$. Changing a representative by a pure-subsector numerator leaves this extension module unchanged. Apply the componentwise denominator and support restrictions of Sec.~\ref{sec:sm-bounded-intersection} directly to this common numerator:

\begin{equation}
\begin{gathered}
\mathcal B_{C,r_0}
 =\mathcal E_{\mathcal Q,r_0}
     \cap V_{\mathcal Q,\boldsymbol d,\mathcal X}.
 \label{eq:sm-bounded-extension}
\end{gathered}
\end{equation}

Here $\mathcal B_{C,r_0}$ is the finite-dimensional numerator space for the off-diagonal row, with $V_{\mathcal Q,\boldsymbol d,\mathcal X}$ now defined on its $N_{\rm sub}$ components. Apply the remainder-matrix construction of Eq.~\eqref{eq:sm-remainder-matrix} using a Gr\"obner basis of $\mathcal E_{\mathcal Q,r_0}$. Dividing the resulting numerator basis by $\mathcal Q$ gives the row ansatz. Its remaining coefficients are determined by the reconstruction conditions of the main text.

\section{Newton-fan strata and generic endpoint regions}
\label{sec:sm-newton-fan}

\subsection{Strata in the space of kinematic weights}
\label{sec:sm-fan-strata}

For two active kinematic scales $x_1,x_2$, introduce a positive asymptotic parameter $u$ and a weight vector $\boldsymbol w=(w_1,w_2)$. Approach infinity along

\begin{equation}
\begin{gathered}
x_1=c_1u^{-w_1},\qquad x_2=c_2u^{-w_2},
 \\
 u\to0,\qquad w_1,w_2\geq0,\quad(w_1,w_2)\neq(0,0).
 \label{eq:sm-infinity-scaling}
\end{gathered}
\end{equation}

The constants $c_1,c_2$ are fixed, nonzero, and generic, so they do not induce additional cancellations in leading coefficients. Positive $w_i$ sends the corresponding scale to infinity; $w_i=0$ holds it fixed. A common positive rescaling of $\boldsymbol w$ only reparametrizes $u$. The relevant Newton fan partitions weight space into cones and their boundary rays according to which polynomial terms form the leading faces. A stratum here is the relative interior of one such cone or ray, where these leading faces remain fixed. In the three benchmark families the strata are represented by

\begin{equation}
\begin{gathered}
\begin{array}{c|c}
 \text{stratum}&\text{representative }(w_1,w_2)\\\hline
 w_1>0,\ w_2=0 &(1,0)\\
 w_1>w_2>0 &(2,1)\\
 w_1=w_2>0 &(1,1)\\
 w_2>w_1>0 &(1,2)\\
 w_2>0,\ w_1=0 &(0,1)
 \end{array}
 \label{eq:sm-newton-strata}
\end{gathered}
\end{equation}

The axes and diagonal are rays; the other representatives lie in the two open cones. The active pairs are $(s,\eta)$ for the bubble and sunset, and $(s,t)$ for the one-external-mass double box. Each stratum must be examined because a hierarchy between scales can expose endpoint regions absent when both scales grow at the same rate. These weights specify the growth of external scales; the Feynman-parameter scalings within each stratum are determined separately below.

\subsection{Endpoint regions from the parametric Newton polytope}
\label{sec:sm-endpoint-regions}

We use the geometric approach to expansion by regions~\cite{Pak:2010pt,Jantzen:2011nz}.

For every contributing sector, let $\boldsymbol\alpha=(\alpha_1,\ldots,\alpha_m)$ be its $m$ Feynman parameters (not to be confused with the pole locations $\alpha$ of the main text), and let $\mathcal U(\boldsymbol\alpha)$ and $\mathcal F(\boldsymbol\alpha;\boldsymbol x)$ be its first and second Symanzik polynomials~\cite{Bogner:2010kv}. At each representative weight, choose an integer $h_{\boldsymbol w}$ large enough to clear the negative powers of $u$ in the kinematically substituted $\mathcal F$. Define the polynomial $\overline{\mathcal F}_{\boldsymbol w}$ and the Newton polytope $\Delta_{\boldsymbol w}$ by

\begin{equation}
\begin{gathered}
\overline{\mathcal F}_{\boldsymbol w}
 =u^{h_{\boldsymbol w}}
    \mathcal F(\boldsymbol\alpha;c_1u^{-w_1},c_2u^{-w_2}),
 \\
 \Delta_{\boldsymbol w}
  =\operatorname{Newt}(\mathcal U\,\overline{\mathcal F}_{\boldsymbol w}).
 \label{eq:sm-endpoint-polytope}
\end{gathered}
\end{equation}

For a polynomial, $\operatorname{Newt}$ means the convex hull of its exponent vectors with nonzero coefficients. Thus each monomial $\boldsymbol\alpha^{\boldsymbol k}u^{\ell_u}$ contributes the point $(k_1,\ldots,k_m,\ell_u)$, where $\boldsymbol k=(k_1,\ldots,k_m)$ records the parameter powers. The polytope is formed in these $m+1$ exponent coordinates, rather than in kinematic space. Multiplication by $u^{h_{\boldsymbol w}}$ translates all points along the last coordinate and leaves the facet normals unchanged.

Compute facets relative to the affine span of the support, the smallest affine space containing all exponent vectors. A facet is a boundary face of codimension one in that space. For each lower facet choose an inward normal $(v_1,\ldots,v_m,v_u)$ with $v_u>0$; ``inward'' means that its scalar product with a support point is minimized on the selected facet. It gives

\begin{equation}
\begin{gathered}
\alpha_i=u^{r_i}\bar\alpha_i,\qquad
 r_i=\frac{v_i}{v_u}-\min_{1\leq j\leq m}\frac{v_j}{v_u}.
 \label{eq:sm-endpoint-scaling}
\end{gathered}
\end{equation}

The exponents $r_i$ specify the region, while the rescaled parameters $\bar\alpha_i$ remain of order one. Under this substitution a monomial has weight $\ell_u+\boldsymbol k\cdot\boldsymbol r$, with $\boldsymbol k\cdot\boldsymbol r=\sum_i k_ir_i$. Terms on the selected facet have the same minimum weight; other terms are suppressed as $u\to0$. Thus the facet specifies which terms balance at leading order.

Feynman parameters are projective: an overall common rescaling does not change their ratios. Subtracting the common minimum in Eq.~\eqref{eq:sm-endpoint-scaling} fixes this freedom by $\min_i r_i=0$. The product $\mathcal U\overline{\mathcal F}_{\boldsymbol w}$ is homogeneous in the Feynman parameters, so the shift does not change its leading face. Facets giving the same projective scaling are identified. The homogeneous region $r_i=0$ is also retained when nonzero, including degenerate cases with no distinct endpoint facet.

For each distinct scaling, substitute into the full parametric integrand, including the measure and propagator powers, extract its overall power of $u$, and expand the remaining factors at fixed $\bar\alpha_i$. Discard scaleless contributions, whose expanded integrals contain no scale. The lowest nonzero powers give the regional leading behavior of the target and masters and depend on their integral indices. When using $\overline{\mathcal F}_{\boldsymbol w}$, restore $\mathcal F=u^{-h_{\boldsymbol w}}\overline{\mathcal F}_{\boldsymbol w}$ with its actual integrand exponent before counting these powers. This procedure enumerates generic endpoint regions; exceptional tangential cancellations and additional Minkowski pinch regions require a separate analysis.

\section{Example families and kinematic settings}
\label{sec:sm-example-details}

\subsection{The bubble families}
\label{sec:sm-bubble-details}

\begin{figure}[htbp]
    \includegraphics[width=6cm]{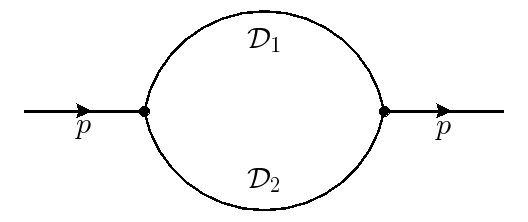}
\caption{\label{fig:sm-bubble}
Feynman diagram of the one-loop bubble families.}
\end{figure}

The single-variable test uses the equal-mass one-loop bubble of Eq.~\eqref{eq:bubble-family}, with $s=p^2$ as the active variable and 2 MIs. The two-variable test uses the unequal-mass bubble defined by $\mathcal D_1=\ell^2-1$, $\mathcal D_2=(\ell+p)^2-\frac14-\eta$, and $p^2=s$, with active variables $(s,\eta)$ and 3 MIs. The common topology is shown in Fig.~\ref{fig:sm-bubble}.

\subsection{The sunset family}
\label{sec:sm-sunset-details}

\begin{figure}[htbp]
%\begin{center}
    \includegraphics[width=6cm]{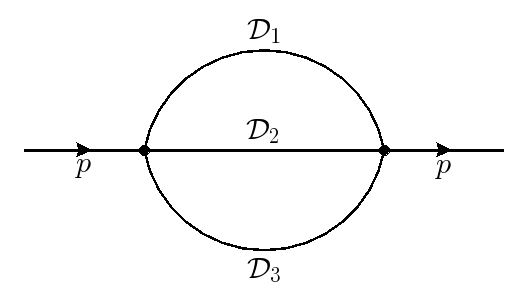}
\caption{\label{fig:sm-sunset}
Feynman diagram of two-loop sunset.}
%\end{center}
\end{figure}

The unequal-mass two-loop sunset is defined by
\begin{align}
 \mathcal{D}_1&=\ell_1^2-1,
 &\mathcal{D}_2&=\ell_2^2-4,
 &\mathcal{D}_3&=(\ell_1+\ell_2+p)^2-9,\nonumber\\
 \mathcal{D}_4&=\ell_1\cdot p,
 &\mathcal{D}_5&=\ell_2\cdot p,
 &p^2&=s,
 \label{eq:unequal-sunset-family}
\end{align}
with 7 MIs; the diagram is shown in Fig.~\ref{fig:sm-sunset}.

The single-variable test uses $s$ as the active variable. For the two-variable test, we replace $\mathcal D_3\to\mathcal D_3-\eta$ and use $(s,\eta)$ as the active variables; the family still has 7 MIs.

\subsection{The planar double-box families}
\label{sec:sm-doublebox-details}

\begin{figure}[htbp]
    \includegraphics[width=6cm]{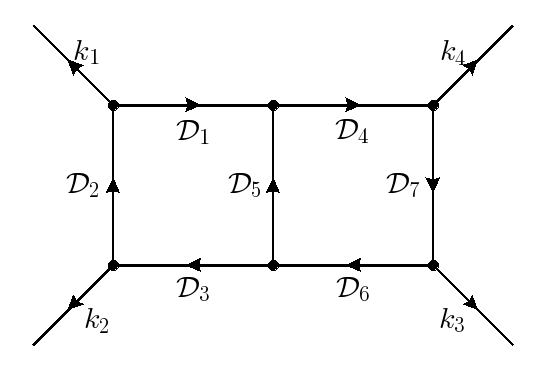}
\caption{\label{fig:sm-doublebox}
Feynman diagram of the planar double-box families.}
\end{figure}

The massless planar double box used in the main text is defined by the seven propagators and two irreducible scalar products
\begin{align}
\begin{split}
    &\mathcal{D}_1={\ell_1}^2,\ \mathcal{D}_2=(\ell_1+k_1)^2,\ \mathcal{D}_3={(\ell_1+k_1+k_2)}^2,\\
    &\mathcal{D}_4= {\ell_2}^2,\ \mathcal{D}_5= {(\ell_2-\ell_1)}^2,\ \mathcal{D}_6={(\ell_2+k_1+k_2)}^2,\\
    &\mathcal{D}_7={(\ell_2-k_4)}^2,\ \mathcal{D}_8 = \ell_1 \cdot k_3,\ \mathcal{D}_9 = \ell_2 \cdot k_1,
\end{split}
\label{eq:massless-doublebox-family}
\end{align}
with $k_i^2=0$, $\sum_{i=1}^4 k_i = 0$; we write $s=(k_1+k_2)^2$ and $t=(k_2+k_3)^2$. The topology is shown in Fig.~\ref{fig:sm-doublebox}.

The auxiliary-mass DE test fixes $s=1$ and $t=5$; the replacement $\mathcal D_1\to\mathcal D_1-\eta$ then introduces $\eta$ as the active variable, and the family has 25 MIs. For the reduction tests, no auxiliary mass is introduced; instead, $k_3^2=1$ while the other external legs remain massless, giving 18 MIs. The single-variable test varies $s$ at fixed $t=3$, whereas the two-variable test varies both $(s,t)$.

\subsection{The nonplanar double-pentagon family}
\label{sec:sm-hjj-details}

\begin{figure}[htbp]
    \includegraphics[width=6cm]{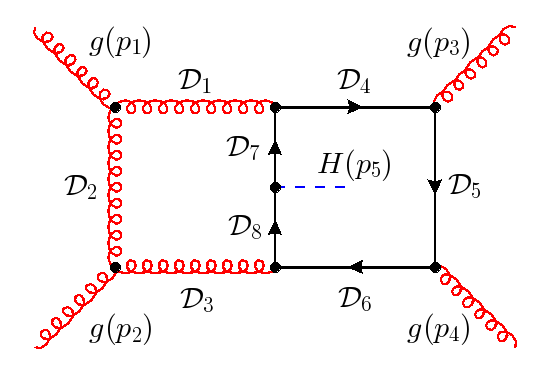}
\caption{\label{fig:sm-hjj}
Topology of the nonplanar double-pentagon family.}
\end{figure}

For the nonplanar double-pentagon family, $p_1+p_2=p_3+p_4+p_5$, $p_i^2=0$ for $i=1,\ldots,4$, and $p_5^2=m_H^2$. With $m_t^2$ as the differential variable, the family is defined by
\begin{align}
 \mathcal D_1&=\ell_1^2,
 &\mathcal D_2&=(\ell_1+p_1)^2,
 \nonumber\\
 \mathcal D_3&=(\ell_1+p_1+p_2)^2,
 &\mathcal D_4&=\ell_2^2-m_t^2,
 \nonumber\\
 \mathcal D_5&=(\ell_2-p_3)^2-m_t^2,
 &\mathcal D_6&=(\ell_2-p_3-p_4)^2-m_t^2,
 \nonumber\\
 \mathcal D_7&=(\ell_1+\ell_2)^2-m_t^2,
 &\mathcal D_8&=(\ell_1+\ell_2+p_5)^2-m_t^2,
 \nonumber\\
 \mathcal D_9&=(\ell_1-p_3)^2,
 &\mathcal D_{10}&=(\ell_1-p_4)^2,
 \nonumber\\
 \mathcal D_{11}&=(\ell_2+p_1)^2.
 \label{eq:hjj-family}
\end{align}
The first eight entries are the physical propagators, and the last three complete the scalar-product basis. Defining $s_{12}=(p_1+p_2)^2$ and $s_{ij}=(p_i-p_j)^2$ for $i=1,2$ and $j=3,4$, we fix
\begin{align}
 s_{12}&=\frac{5}{2},&
 s_{13}&=-\frac{4}{3},&
 s_{14}&=-\frac{7}{4},
 \nonumber\\
 s_{23}&=-\frac{9}{5},&
 s_{24}&=-\frac{11}{6},&
 m_H^2&=1.
 \label{eq:hjj-kinematic-point}
\end{align}
The topology is shown in Fig.~\ref{fig:sm-hjj}.

In the {\tt AMFlow} comparison of the main text, we use {\tt RecursionMode $\to$ "AMF"} and {\tt AMFMode $\to$ "Loop"}: the auxiliary mass $\eta$ is inserted into $\mathcal D_4,\ldots,\mathcal D_8$, and the resulting family coincides with the one above upon identifying $\eta=m_t^2$. In Table~\ref{tab:hjj-amflow-timing}, the {\tt Blade}+{\tt AMFlow} row is the measured result with {\tt Blade} as the IBP reducer and the {\tt AMFlow}~2.0 {\tt C++} differential-equation solver, while the AB+{\tt Blade}+{\tt AMFlow} row scales the construction cost by the IBP-point reduction $5/372$ per prime, leaving the solution time unchanged.

\section{Asymptotic expansion results}
\label{sec:sm-regional-details}

In each region $r$ with local expansion variable $z$, expanding the current-sector side of the top-sector DE, $\partial_\eta\vec I_{\rm top}-A_{11}(\eta)\vec I_{\rm top}=A_{12}(\eta)\vec I_{\rm sub}$, and every subsector master, followed by regional IBP reduction, gives
\begin{align}
 \left.\left(\partial_\eta \vec{I}_{\rm top}
 -A_{11}(\eta)\vec{I}_{\rm top}\right)\right|_r
 &=\sum_{\gamma,p,\nu}
  \vec{X}^{(r)}_{\gamma p\nu}\,
  z^\gamma(\log z)^p \mathcal J_\nu^{(r)},
  \nonumber\\
 \left.\vec{I}_{\mathrm{sub}}\right|_r
 &=\sum_{\gamma,p,\nu}
  \vec{Y}^{(r)}_{\gamma p\nu}\,
  z^\gamma(\log z)^p \mathcal J_\nu^{(r)} ,
  \label{eq:regional-master-expansions}
\end{align}
where $\mathcal J_\nu^{(r)}$ are the MIs of the regional topology, and the coefficients $\vec X^{(r)}$ and $\vec Y^{(r)}$ are obtained from the direct regional expansion and its IBP reductions.

Table~\ref{tab:top-regional-reconstruction} identifies the momentum scaling of every region and the constraint rank from asymptotic expansions for the double-box top sector of the main text. The two top-sector DE rows are treated jointly. The two collinear regions at $\eta=0$ are treated as one overlap envelope.  The column ``regional MIs'' is the number of $\mathcal J_\nu^{(r)}$.   The last column gives the joint constraint rank supplied by the leading matching equations of the indicated region; the final row is the rank obtained by stacking the equations from all listed regions.
\begin{table*}[ht]
\caption{Regional topologies and leading-order constraint ranks used to determine the two top-sector off-diagonal rows.}
\label{tab:top-regional-reconstruction}
\begin{ruledtabular}
\begin{tabular}{llcc}
endpoint&momentum scaling / regional topology&regional MIs&constraint rank\\
$\eta=0$&$\ell_1$ $k_1$-collinear, $\ell_2$ hard&5&8\\
$\eta=0$&both loops $k_1$ or $k_4$-collinear&4&3\\
$\eta=1$&$\ell_1+P$ soft ($P=k_1+k_2$), $\ell_2$ hard&1&2\\
$\eta=1$&$\ell_1+P$ ultrasoft, $\ell_2+P$ $k_3$-collinear&2&2\\
$\eta=1$&$\ell_1+P$ and $\ell_2+P$ soft&2&4\\
$\eta=\infty$&all-hard&1&2\\
$\eta=\infty$&$\ell_1$ hard, $\ell_2$ soft&1&2\\
$\eta=\infty$&all-soft&5&6\\
\hline
all listed endpoints&cumulative system&--&22\\
\end{tabular}
\end{ruledtabular}
\end{table*}

\section{Choice of master integrals for differential equations}
\label{sec:sm-basis-details}

The effect of the master basis on the factorization of IBP denominators has been studied in Refs.~\cite{Smirnov:2020quc,Usovitsch:2020jrk,Boehm:2020ijp}. For integral reduction, an invertible change of MI basis transforms the reduction coefficients and their numerator module together. If the denominator and degree-cap inputs are transformed consistently, the bounded module ansatz is therefore mainly reparametrized, and its dimension is usually only mildly affected.  This relative insensitivity is useful in practice: the reduction ansatz can be constructed in a convenient master basis without a separate basis-optimization step.

The situation changes when bootstrapping the DEs.  The pole structure of each diagonal block then determines the extensions for the corresponding off-diagonal blocks, so a change of basis can alter the number of extension directions substantially. A spurious simple pole $1/(x-\alpha)$ introduces a new analyticity condition of order $\mathcal{O}(x-\alpha)$.  At an existing singularity, a pole of order $r>1$ also introduces a new analyticity condition of order $\mathcal{O}((x-\alpha)^r)$. These changes can enlarge the diagonal source module and hence the space of extension-compatible directions for the off-diagonal blocks.

We demonstrate the effect with two top-sector bases of the massless double box of the auxiliary-mass DE example in the main text.  Replacing $I_{1,1,1,1,1,1,1,0,0}$ by $I_{1,1,2,1,1,1,1,0,0}$, with $\mathcal T$ the corresponding basis-change matrix, gives
\begin{align}
 \det\mathcal T=\frac{9(\eta+1)}{7(\eta-1)^2}.
\end{align}
At the chosen kinematics the finite Landau roots are $\eta=0,1,-5,5/6$; hence $\eta=-1$ is a spurious divisor. Table~\ref{tab:spurious-basis-comparison} uses the column conventions of Table~\ref{tab:de-bootstrap-summary}. Here the dimensions refer to the first top-sector off-diagonal row, and the module-extension comparison is made before imposing the infinity constraints. ``Top-sector basis'' identifies the basis replacement. The spurious divisor enlarges the extension-compatible ansatz from 11 to 27 parameters.  This is the basis dependence relevant to DE bootstrap: poles in a diagonal block create additional local directions that must be extended into its subsectors.

\begin{table}[htbp]
\caption{Dependence of the top-sector off-diagonal ansatz on the master-integral basis. Dimensions are counted before imposing the infinity constraints; after imposing them, the Fuchsian-basis dimension reduces from 11 to 10, as quoted in the main text.}
\label{tab:spurious-basis-comparison}
\begin{ruledtabular}
\begin{tabular}{lrr}
top-sector basis&trivial dimension& AB dimension\\
Fuchsian basis&30&11\\
transformed basis&79&27\\
\end{tabular}
\end{ruledtabular}
\end{table}

\section{Integral-reduction tests}
\label{sec:sm-reduction-tests}

The benchmark families and kinematic settings are defined in Sec.~\ref{sec:sm-example-details}. The following tables give the numerator-ansatz dimensions and finite-field reconstruction costs.

\subsection{Single-variable reductions}
\label{sec:sm-single-variable-details}

The single-variable tests scan dots $2$--$10$, $2$--$8$, and $2$--$6$ for the equal-mass bubble, unequal-mass sunset, and one-external-mass planar double box, respectively.

Table~\ref{tab:single-variable-reductions} shows the results of the numerator module ansatz reconstruction for the three single-variable families, following the column conventions defined with Table~\ref{tab:de-bootstrap-summary}.
\begin{table*}[t]
\caption{Single-variable reduction ansatzes and their finite-field reconstruction costs. }
\label{tab:single-variable-reductions}
\begin{ruledtabular}
\begin{tabular}{lcrrrrr}
family&number of MIs&target&trivial dimension&AB dimension&trivial points&AB points\\
equal-mass bubble&2&$I_{2,2}$&4&1&2&1\\
&&$I_{2,3}$&5&1&3&1\\
&&$I_{3,3}$&7&1&4&1\\
&&$I_{3,4}$&8&1&5&1\\
&&$I_{4,4}$&10&1&6&1\\
&&$I_{4,5}$&11&1&7&1\\
&&$I_{5,5}$&13&1&8&1\\
&&$I_{5,6}$&14&1&9&1\\
&&$I_{6,6}$&16&1&10&1\\
\hline
unequal-mass sunset&7&$I_{1,2,2,0,0}$&23&3&4&1\\
&&$I_{2,2,2,0,0}$&39&1&6&1\\
&&$I_{2,2,3,0,0}$&65&5&10&1\\
&&$I_{2,3,3,0,0}$&82&4&13&1\\
&&$I_{3,3,3,0,0}$&106&7&16&1\\
&&$I_{3,3,4,0,0}$&124&6&19&1\\
&&$I_{3,4,4,0,0}$&148&9&23&2\\
\hline
one-mass double box&18&$I_{1,1,1,3,1,1,1,0,0}$&69&21&6&2\\
&&$I_{1,1,1,4,1,1,1,0,0}$&99&32&8&2\\
&&$I_{1,1,1,5,1,1,1,0,0}$&129&43&10&3\\
&&$I_{1,1,1,6,1,1,1,0,0}$&159&54&12&4\\
&&$I_{1,1,1,7,1,1,1,0,0}$&189&65&14&4\\
\end{tabular}
\end{ruledtabular}
\end{table*}

\subsection{Two-variable reductions}
\label{sec:sm-multivariable-details}

The two-variable tests scan dots $2$--$8$, $2$--$8$, and $2$--$6$ for the unequal-mass bubble, unequal-mass sunset, and one-external-mass double box, respectively.

Table~\ref{tab:multivariable-reductions} shows the results of the numerator module ansatz reconstruction for the three two-variable families of the main text, following the column conventions defined with Table~\ref{tab:de-bootstrap-summary}.
\begin{table*}[t]
\caption{Multivariable reduction ansatzes and finite-field reconstruction costs.}
\label{tab:multivariable-reductions}
\begin{ruledtabular}
\begin{tabular}{lcrrrrr}
family&number of MIs&target&trivial dimension&AB dimension&trivial points&AB points\\
unequal-mass bubble&3&$I_{2,2}$&18&1&6&1\\
&&$I_{2,3}$&35&1&15&1\\
&&$I_{3,3}$&57&1&21&1\\
&&$I_{3,4}$&85&1&36&1\\
&&$I_{4,4}$&118&1&45&1\\
&&$I_{4,5}$&157&1&66&1\\
&&$I_{5,5}$&201&1&78&1\\
\hline
unequal-mass sunset&7&$I_{1,2,2,0,0}$&54&1&10&1\\
&&$I_{2,2,2,0,0}$&189&1&28&1\\
&&$I_{2,2,3,0,0}$&451&1&66&1\\
&&$I_{2,3,3,0,0}$&795&1&120&1\\
&&$I_{3,3,3,0,0}$&1198&1&190&1\\
&&$I_{3,3,4,0,0}$&1772&1&276&1\\
&&$I_{3,4,4,0,0}$&2406&1&378&1\\
\hline
one-mass double box&18&$I_{1,1,1,3,1,1,1,0,0}$&269&8&28&1\\
&&$I_{1,1,1,4,1,1,1,0,0}$&495&18&45&2\\
&&$I_{1,1,1,5,1,1,1,0,0}$&793&29&66&3\\
&&$I_{1,1,1,6,1,1,1,0,0}$&1163&43&91&3\\
&&$I_{1,1,1,7,1,1,1,0,0}$&1605&62&120&4\\
\end{tabular}
\end{ruledtabular}
\end{table*}

%%%%%%%%%%%%%%%%%%%%%%%%%%%%%%%%%%%%%%%%%%%%%%%%%

\end{document}